# Performance Evaluation of Spread Spectrum Watermarking using Error Control Coding

T. S. Das[1], V. H. Mankar[2], S. K. Sarkar[3]
Jadavpur University, Kolkata-700032 INDIA
[1]tirthasankardas@yahoo.com, [2]vijaymankar@yahoo.com, [3]sksarkar@etce.jdvu.ac.in

*Abstract*-This paper proposes an oblivious watermarking algorithm with blind detection approach for high volume data hiding in image signals. We present a detection reliable signal adaptive embedding scheme for multiple messages in selective sub-bands of wavelet (DWT) coefficients using direct sequence spread spectrum (DS-SS) modulation technique. Here the impact of volumetric distortion sources is analyzed on the ability of analytical bounds in order to recover the watermark messages. In this context, the joint source-channel coding scheme has been employed to obtain the better control of the system robustness. This structure prevents the desynchronisation between encoder and decoder due to selective embedding. The experimental results obtained for Spread Spectrum (SS) transformed domain watermarking demonstrate the efficiency of the proposed system. This algorithmic architecture utilizes the existing allocated bandwidth in the data transmission channel in a more efficient manner.

*Keywords*- Spread Spectrum, DWT, Source & Channel coding, Signal Adaptive, Blind Detection.

## I. INTRODUCTION

The vast progress of digital technologies has contributed to popularize the use of electronic media for transmission and storage of multimedia information. Data stored in digital format can be copied without loss of quality and distributed efficiently at fairly low cost. These developments have also greatly enhanced the potential for interception, manipulation and unauthorized distribution of information. For this reason, the design of efficient techniques for preserving the ownership of digital content is the fundamental issue to be addressed for the future multimedia services [1].

The DS-SS modulation technique in digital communication offers anti-jamming and interference rejection property. In DS-SS communication, a low-level wide band signal can easily be hidden within the same spectrum as a high power signal where each signal appears as a noise to the others. At the receiver, low-level wide band signal will be accompanied by noise, and by using a suitable detector this signal can be squeezed back into the original narrow base-band. Because noise is completely random and uncorrelated the wanted signal can easily be extracted. These had motivated the several researchers for developing SS watermarking algorithms for multimedia signals either in spatial domain or in transform domain by using DCT, Fourier-Mellin, DHT, and wavelet decomposition [2]. DS-SS watermarking schemes, although can be implemented in various different ways; the method that uses distinct pseudo noise (PN) spreading codes for embedding each binary digit is popular and proven to be efficient, robust and cryptographically secure. At this point the use of various source and channel-coding or *M*-ary modulation schemes can be found efficient for robustness improvement as they are widely used in digital communication for increasing data transmission reliability [3-4].

Based on this concept we may construct an encoding scheme for conveying message or watermark with some extra information bits via a transmission channel. This extra information will be added into the original signal by using source-channel coding framework before transmission process takes place. At the receiver this extra information helps to detect the watermark properly thereby providing better objective as well as subjective recognition.

In the applications of digital image watermarking the concept of *M*-ary modulation also provides moderate results regarding subjective recognition of the embedded data. With few numbers of modulation function and large modulation index, the *M*-ary scheme becomes inefficient under burst error condition. This motivates us to investigate in the field of source and channel coding in order to improve the watermarking system robustness.

In this context, joint source-channel coding framework become efficient with respect to better subjective recognition of the hidden watermark even if it undergoes severe image impairments. In source-channel coding framework, computational cost and complexity concerning PN sequence generation is much reduced and simple correlator can be used as a detector structure instead of complex maximum likelihood receiver.

The paper is organized as follows: Section II introduces proposed SS watermarking and detection. Error Control Coding (ECC) is given in Section III. Section IV presents the watermarking architecture implemented in the present work. Section V shows the experimental results on 16-by-16 binary watermark and finally section VI concludes and remarks about some of the aspects analyzed in this paper.

## II. SS WATERMARKING AND DETECTION

Consider $C_0$ a vector of length $M$ containing the elements of the host image. We want to embed into $C_0$ a multi-bit message $B$ of length $N$.

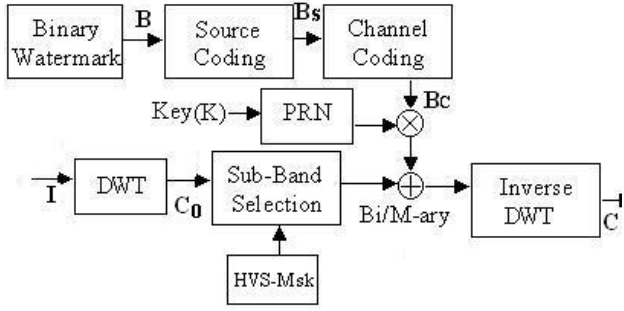

Fig. 1. Encoder

$$B = [B_1, B_2, ..., B_N], \quad B_j \varepsilon \{-1,+1\} \quad (1)$$

To obtain the watermark vector $W$ of length $M$ carrying the $N$-bit message, we construct a set $P$ of $N$ reference marks. Each reference mark $P_j$ is a pseudo-random sequence of length $M$.

$$P = [P_1, P_2, ..., P_N] \quad (2)$$
$$P_j = [P_{j1}, P_{j2}, ..., P_{jM}], \quad P_{ji} \varepsilon \{-1,+1\} \quad (3)$$

The pseudo-random sequence elements $P_{ji}$ are random numbers assuming −1 or +1 according to a uniform, zero-mean probability density function. For DS-SS system, the pseudo-random sequences of the set $P$ should be orthogonal to each other. The initial state of the PN sequence generator defines the cryptographic key $K$ of the watermarking system. This key $K$ should be known by both embedder and detector such that they generate the same set $P$. Non-orthogonality between reference marks results in a sort of inter-symbol interference that may compromise the embedding efficiency of the system. In the following equations we will show how to achieve total embedding efficiency. By equation (4) we spread the $N$-bit message $B$ into an $M$-dimensional sequence $W$ corresponding to the watermark vector. The gain factor $\alpha$ determines the watermark magnitude.

$$W = \alpha \sum_{j=1}^{N} B_j \, P_j \quad (4)$$

In a blind embedding scheme, we obtain the watermarked image vector by $C_W = C_0 + W$. Blind embedding schemes do not assure embedding efficiency and also do not explore properties of the HVS to improve the perceptual transparency [13-14]. To introduce perceptual modeling improving watermark transparency, we propose the use of a multiplicative mask $Msk$ according to (5).

$$C_W = C_o + Msk * W \quad (5)$$

The operation * is defined for the element-by-element multiplication of the vectors. Vector $Msk$ is a masking image with elements ranging from 0 to $msk_{max}$ according to the insensitivity of the perceptual impact of the watermark added to the original image. The mask is obtained by the squared root of the absolute value of the elements of the Inverse DiscreteWavelet Transform (IDWT) image reconstruction considering only the detail sub-bands (*HL*1, *LH*1 and *HH*1). Note that regions of edges and texture will receive more watermark energy. To assure total embedding efficiency, we need to propose an embedding scheme that compensates the non-orthogonalities between host image and reference pattern vectors. Considering the use of linear correlation for message recovery at decision variable $Dk$ of equation (6).

$$D_k = \langle C_W, P_k \rangle = \frac{1}{M} \sum_{i=1}^{M} C_{Wi} P_{ki}$$
$$= \left\langle P_k, \left[ C_o + \alpha . \sum_{i=1}^{N} B_j Msk * P_j \right] \right\rangle$$
$$= \langle P_k, C_0 \rangle + \alpha . B_k \langle P_k, Msk * P_k \rangle + \alpha . \sum_{i=1, j \neq k}^{N} B_j \langle P_k, Msk * P_j \rangle \quad (6)$$

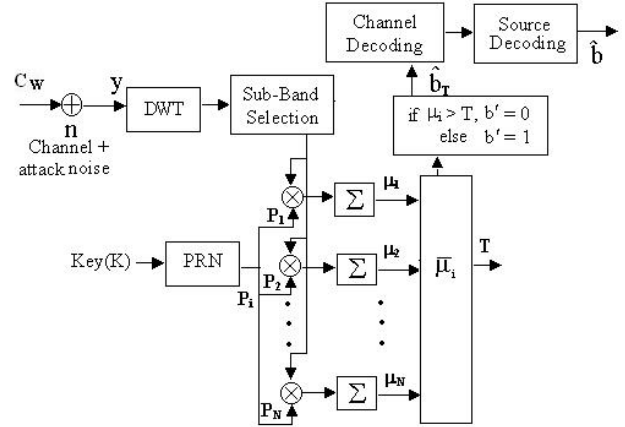

Fig. 2. Decoder

For the $N$-bit message recovery at detector, we have to evaluate $D_k$ and take a binary decision for $k = 1, 2, ..., N$. If $D_k > 0$, $B_k = +1$, otherwise $B_k = −1$. Since $P_{ki} \varepsilon \{−1, +1\}$, we know that $\langle P_k, Msk * P_k \rangle$

$$= \frac{1}{M} \sum_{i=1}^{M} Msk_i \, P_{ki}^2 = \mu_{Msk}$$

For simplification, if we call $R_k^{C0} = \langle P_k, C_0 \rangle$ and

$$R_k^{(Msk*P_j)} = \alpha . \sum_{i=1, j \neq k}^{N} B_j \langle P_k, Msk * P_j \rangle, \text{ then equation (6) results in}$$

$$Dk = \alpha \cdot B_k \cdot \mu_{Msk} + (R_k^{C0} + R_k^{(Msk*P_j)}) \quad (7)$$

Unfortunately, $R_k^{C0} = 0$ and $R_k^{(Msk*P_j)} = 0$ cannot be assumed. These residual components from non-orthogonalities may cause bit errors. To assure the system embedding efficiency, we need to compensate the harmful interference of these residual components in the decision variable $Dk$ setting an appropriate $\alpha$ value. As we have different residual components for each decision variable $Dk$, we make the following change in the step of watermark construction (8) to allow the appropriate setting of individual gain factors $\alpha_j$ for each reference mark.

$$W^* = \sum_{j=1}^{N} \alpha_j B_j \, P_j \quad (8)$$

Generalizing the existence of volumetric distortion sources in the watermarked image path, we define a distortion vector $n$ that changes the elements of the watermarked image vector $C_W$ according to (9).

$$C_W' = C_W + n \quad (9)$$

Finally, performing the watermark construction and embedding steps according to equations (8) and (5) we obtain the proposed multi-bit watermarking system (10) that assures total embedding efficiency.

$$C_W = C_0 + Msk * W^* \quad (10)$$

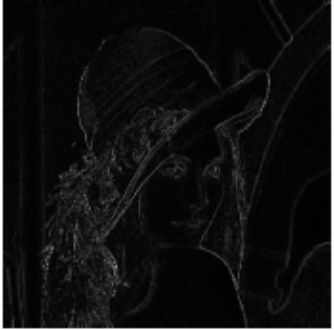

Fig. 1 Wavelet based mask for watermark embedding

### III. ERROR CONTROL CODING

#### A. Golomb Code as Source coding:

As we said in the last section, the sequence of predictions consists of (usually long) segments of zeros separated by (usually short) segments of ones. An efficient method to compress this type of information is the Golomb code. Some other methods based on the Golomb code (as LOCO-I, FELICS and JPEG-LS) also seem to be efficient, however we did not test them [5].

The Golomb code is used to encode sequences of zeros and ones, where a zero occurs with (high) probability $p$ and a one occurs with (low) probability 1-$p$. The Golomb code depends on the choice of an integer parameter $m \geq 2$ and it becomes the best prefix code when

$$m = \left[-\frac{\log_2(1+P)}{\log_2 P}\right] \quad (11)$$

For small values of $m$, the Golomb codes start short and increase quickly in length. For large values of $m$, the Golomb codes start long, but their lengths increase slowly. To compute the code of a nonnegative integer $n$, three quantities $q$, $r$ and $c$ are computed:

$$q = \frac{n}{m}, r = n - qm, c = \log_2 m \quad (12)$$

Then, the code is constructed in two parts: the first is the value of $q$, coded in unary, and the second is the binary value of $r$ coded in a special way. If $r<2c-m$, $r$ is coded as unsigned integers in $c$-1 bits. If $r \geq 2c-m$, $r$ is represented as the unsigned integer $r+2c-m$ in $c$ bits. The case where $m$ is a power of 2 is special because it requires no ($c$-1)-bit codes (called Rice codes). To decode a Golomb code, the values of $q$ and $r$ are used to reconstruct $n$ ($n=r+qm$).

#### B. Convolution Code as Channel Coding:

Since the image restoration does not result in a perfect copy of the original cover image and the embedded signal is low power, the estimate of the embedded signal is poor. This results in a demodulated message signal that may have a substantial number of bit errors, indicated by a high-embedded signal BER (typically greater than 0.15 BER). Therefore, to allow for the sub optimal performance of the signal estimation process, we have incorporated the use of low-rate error-control codes to correct the large number of bit errors. Any error-correcting code that is capable of correcting the high signal estimation BER can be used within SS. For SS proof-of-concept, binary expansions of Reed–Solomon codes with a decoder based on a simple idea of Bossert and Hergert using low-weight parity checks were used for error correction. This has now been extended to convolutional codes using the Viterbi algorithm [6]. The use of error correction by DS-SS compensates for the sub optimal estimation of the embedded signal, in addition to combating distortion, which may be encountered during transmission of the stegoimage. At the encoder, the entire decoding process can be simulated, thus permitting selection of the proper error-correcting code for the chosen cover image and embedded signal strength. This allows the assurance that the hidden message can be recovered, with high probability and error free when the transmission channel is noiseless. When the transmission channel is expected to be noisy, it is possible that an appropriate error correcting code may be selected to correct for the additional errors caused by the channel. Similarly, the error correction may be able to compensate for the errors generated by the use of low levels of compression applied to the stego image.

### IV. WATERMARKING ARCHITECTURE

The proposed work considers a binary image of size (16 X 16) as watermark and (256 X 256), 8 bits/pixel gray image as host/ cover image. In order to show the better robustness of proposed scheme data is source and channel encoded by joint Golomb-convolution coding techniques. Now, the question is what features of cover signal is suitable for robust watermarking? There is probably no answer to this question as different features have different levels of robustness to certain attacks. Wavelet transform domain provides important class of features for data hiding as well as the co-joint representation of simultaneous space-frequency resolution of image signal. Wavelet coefficients are more efficient in representing perceptually important signal features and thus potentially more robust to distortion. Wavelet transform also attracts attention in various image processing applications including de-noising and upcoming compression standard JPEG-2000 due to its specific level of robustness. The host image can be modeled as approximately i.i.d., sequence with gaussian distribution by this wavelet transform. Here linear, additive modulation function is used for data embedding. Therefore, all watermark features are treated equally and spread over them evenly. Now, in order to accomplish better spectrum spreading data is embedded in *LL* and *HH* sub bands of DWT decomposition while the same is done into few selected channels with low and high variance value in *M*-Band WT

*Performance Evaluation of Spread Spectrum Watermarking using Error Control Coding*

decomposition domain. A single or a set of binary valued PN sequence equal to the size of sub band/ channel are generated and for each PN matrix, the orthogonal code is obtained by complementing the bits of PN code. If PN code is used for data embedding in *LL* sub band ($H_{12}$, $H_{13}$, $H_{14}$, $H_{24}$ in *M*-Band), the orthogonal code $\overline{(PN)}$ is used for data embedding in *HH* sub band ($H_{41}$, $H_{42}$, $H_{43}$, $H_{31}$) [7-9].

## V. RESULTS AND DISCUSSION

The Spread Spectrum (SS) watermarking scheme is applied in wavelet transform domain over large number of benchmark images. It is quite clear that robustness efficiency is improved to a greater extent under the burst error condition with the increase of payload amount. At the same time computational cost and complexity regarding code pattern, decoding and subjective recognition are significantly improved. The proposed scheme performs well over the conventional antipodal binary and *M*-ary scheme because of its inherent error correction capability. *M*-ary scheme provides good results with increasing amount of *M* but it seems to be impractical during decoding when *M* is very large ($M > 256$). Under this constraint scenario, channel coding also increases the overhead of redundant extra bits though it performs well in respect to error correction during decoding. This reduces the overall efficiency of the channel coding schemes. Therefore source-coding mechanisms are incorporated in the architecture, which compensates the increased overhead added by the channel coding techniques. The source coding removes the redundancy from the watermark message and thus the error correction capability can be effectively utilized without increasing the effective transmission bandwidth. In this way the joint Golomb-Convolution coding framework minimize the residual effect between the image and PN sequences thereby detection reliability will be highly ensured. The experimental results are given in fig.1 and Table I.

## VI. CONCLUSION

In this paper we have studied the application of joint source - channel coding structure in wavelet transformed SS watermarking system for multimedia signals specifically gray scale images. Since DWT approximates the host image distribution to a gaussian one, therefore correlator is used as the optimal blind detector. The detection reliability provides an analytical indicator for BER (bit error rate), which can be used to know the achievable performance. We observe that the use of Golomb and convolution code combination results in a significant improvement of the BER against several volumetric impairments. The main factors that determine the difference in performance are the minimum distance and the redundancy of the codes.

TABLE I
NUMERICAL RESULTS

| Image | Filter | SSIM | PSNR | Security Value |
|---|---|---|---|---|
| Fishing Boat | DWT/ db2 | 0.973 | 36.18 | 0.0210 |
| | M Band | 0.975 | 38.041 | 0.0162 |
| Lena | DWT/ db2 | 0.960 | 37.31 | 0.0197 |
| | M Band | 0.972 | 38.45 | 0.0160 |

TABLE II
JPEG -2000 COMPRESSION

| QF | BER Without ECC | BER With ECC |
|---|---|---|
| 100 | 0.26 | 0 |
| 75 | 0.32 | 0 |
| 50 | 0.46 | 0 |
| 35 | 0.57 | 0 |
| 25 | 0.66 | 0.27 |
| 5 | 0.78 | 0.38 |

[Results of Table II are obtained using watermark length 256, source (Golomb) & channel (Convolution) coded watermark length 404]

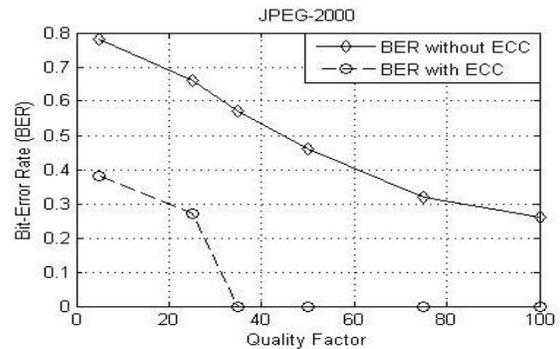

Fig. 1. Compression results are obtained using original watermark length 256, source (Golomb) & channel (Convolution) coded watermark length 404